\documentclass[a4paper,
               biblatex,     
               keeplastbox,   
               ]{jacow}
\usepackage{pdfpages,multirow,ragged2e} %
\makeatletter%
	\ifboolexpr{bool{xetex}}
	 {\renewcommand{\Gin@extensions}{.pdf,%
	                    .png,.jpg,.bmp,.pict,.tif,.psd,.mac,.sga,.tga,.gif,%
	                    .eps,.ps,%
	                    }}{}
\makeatother

\ifboolexpr{bool{xetex} or bool{luatex}} 
 {}                                      
 {\usepackage[utf8]{inputenc}}           

\usepackage[USenglish]{babel}
\usepackage{nameref}

\begin{document}

\title{From SKA to SKAO: Early Progress in the SKAO Construction}

\author{J. Santander-Vela\textsuperscript{1}\thanks{juande.santander-vela@skao.int},  M. Bartolini\textsuperscript{1,2}, M. Miccolis\textsuperscript{1}, N. Rees\textsuperscript{1}\\
	\textsuperscript{1}SKA Observatory, SK11 9FT Jodrell Bank, United Kingdom\\
	\textsuperscript{2}INAF Istituto Nazionale di Astrofisica, Viale del Parco Mellini 84, 00136 Roma, Italy}
	
\maketitle

\begin{abstract}
   The Square Kilometre Array telescopes have recently started their 
   construction phase, after years of pre-construction effort. The new 
   SKA Observatory (SKAO) intergovernmental organisation has been 
   created, and the start of construction ($\mathrm{T_{0}}$) has 
   already happened. In this talk, we summarise the construction 
   progress of our facility, and the role that agile software 
   development and open-source collaboration, and in particular the 
   development of our TANGO-based control system, is playing.
\end{abstract}

\section{Introduction} 
\label{sec:introduction}

The Square Kilometre Array (SKA) is an international project that has the aim of building two multi-purpose radio telescope arrays. One of them will be built in South Africa in the Karoo desert, and the other will be constructed in the Murchison Shire in Western Australia. The name comes from the initial intention for these telescopes to provide the equivalent collecting area of at least one square kilometre, and thus unprecedented sensitivity, which would allow key questions in modern astrophysics and cosmology to be answered.

The original \emph{Hydrogen Array} concept of an array that was sensitive enough through a very large collecting area of up to one square kilometre was described by Peter Wilkinson in 1991~\cite{1991ASPC...19..428W}. One of the main concepts took the name of the Square Kilometre Array project, and several milestones were achieved in order to make this project a reality.

After several forms (from an interest group to the International SKA Project Office, later the SKA Project Office in Manchester University), and several EU framework programs (SKADS, the SKA Design Study; PrepSKA, preparation for SKA), the SKA Organisation was founded in November 2011 as a non-for-profit limited responsibility company established in England and Wales.

As part of SKADS, the first  SKA Science book was published in 2004~\cite{2004NewAR..48..979C}. After the official start of the SKA Pre-Construction in 2014, an update to the SKA science book was published~\cite{2015aska.confE.....} in 2015 after a decade of development of the SKA concept, incorporating more than 130 scientific use cases that will be possible thanks to the SKA telescopes.

Those science cases cover Galaxy Evolution, Cosmology and Dark Energy\footnote{http://skatelescope.org/galaxyevolution/},  Strong-Field Tests of Gravity\footnote{http://skatelescope.org/gravity-einstein-pulsars/}, Cosmic Magnetism\footnote{http://skatelescope.org/magnetism/}, The Cosmic Dawn and the Epoch of Reionisation\footnote{http://skatelescope.org/cosmicdawn/}, and research on the Cradle of Life\footnote{http://skatelescope.org/cradle-life/}. The amount of physical disciplines foreseen to be encompassed by the SKA telescopes is one of the largest for any ground based facility to date.

The SKA project is currently in what is known as SKA Phase 1, or SKA1, in which two telescopes approximately with 10\% of the target collecting area are being built, namely SKA1-Mid, and SKA1-Low, in order to prove the feasibility of the techniques and derisk the construction of the next phase of the project, SKA Phase 2, or SKA2.

The goal is to have a single observatory entity, that will construct and operate two SKA1 telescopes (SKA1-Mid and SKA1-Low), with presence in three sites: Australia (SKA1-Low), South Africa (SKA1-Mid), and United Kingdom (Headquarters and central operations).

This talk focuses on the progress and status of the SKA project from our last status report~\cite{Santander-Vela:ICALEPCS2017-FRAPL01} in ICALECPS'17. It starts by describing how we have migrated from the SKA Organisation and pre-construction towards the SKA Observatory (SKAO) in~\textsc{\nameref{sec:ska_to_skao}}. We later indicate the role of software in the SKA project in~\textsc{\nameref{sec:sw_in_ska}}, and we provide an update on the status of our efforts in~\textsc{\nameref{sec:current_status}}. We  continue by describing the difficulties that we have been facing up to the start of construction in~\textsc{\nameref{sec:challenges}}, and we describe future work in~\textsc{\nameref{sec:next_steps}}, with some short~\textsc{\nameref{sec:conclusion}} at the end.


\section{From SKA to SKAO and Start of Construction} 
\label{sec:ska_to_skao}

As indicated in Sec.~\nameref{sec:introduction}, the Hydrogen Array concept was first published in 1991. Several studies were made to come with a concept for the realisation of that square kilometre array, and in 2008 the EU Framework Programme called SKA Design Studies (SKADS) was started, and then followed by PrepSKA. After PrepSKA, the SKA Organisation was founded as a non-for-profit, limited liability company registered in England and Wales, but it also set in motion the process for finding what would be the ultimate legal form for the SKA Observatory (SKAO), and it was finally decided that an Inter-Governmental Organisation (IGO) was the way to go.

After PrepSKA, which set up the first contacts towards the IGO, the first round of negotiations towards the SKAO IGO took place in October 2015. Several rounds of negotiations where required, and they culminated with the signature of the SKAO Treaty in Rome in March 2019.

For the treaty to enter into force, it was required that the three site countries (UK, Australia, and South Africa) had to ratify the treaty, plus two more countries. The process was finally complete with the ratification of the Treaty in December 2020, and the entry into force of the Treaty happened in January 15th, 2021.

After that, the first SKAO Council met on February 4th, 2021, setting up the way for the acquisition of the assets of the SKA Organination, with personal transferred through the TUPE process. All company assets and staff were transfered to the IGO in May 1st, 2021.

\begin{figure*}[tb]
  \centering
    \includegraphics[width=\textwidth]{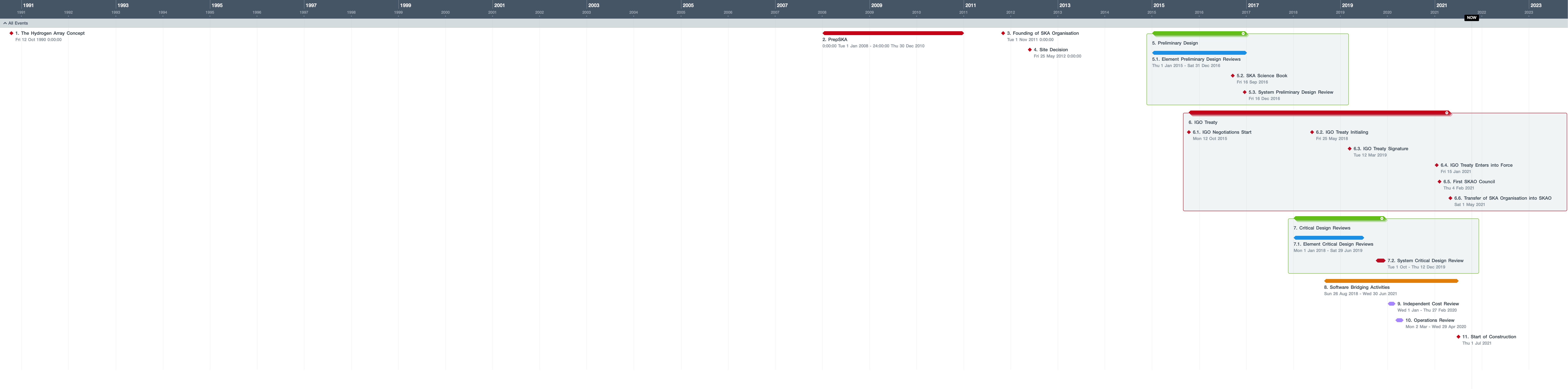}
  \caption{Timeline of the SKA project pre-construction phase. The earliest milestone took place in 1991 with the first concept of \emph{The Hydrogen Array}. The boxes in green represent the design phases, while the items in read represent progress towards the establishment of the SKA project and the SKAO IGO. The final date for the pre-construction phase is $T_0$, the Start of Construction.}
  \label{fig:ska-timeline}
\end{figure*}

All of this process can be seen in detail in Fig.~\ref{fig:ska-timeline}.

In June 2021 the SKA1 Construction Proposal (CP), and the SKA Observatory Establishment and Delivery Plan (OEDP) were submitted by the SKAO Director-General with a recommendation for approval by the Council. The CP describes the science requirements for the project; what the scope of the project is; how the project will be executed, monitored, and controlled; wider benefits of the project to society; and a guide to the detailed project documentation, reference information, and its organisation; whereas the OEDP details staffing and costs for the SKA Observatory own delivery and supporting functions: Business enabling functions, Observatory operations, Observatory development, and Construction support. The Council approved both documents, and hence the start of construction ($T_0$) took place on July 1st, 2021.


\section{Software in the SKA Project} 
\label{sec:sw_in_ska}

\begin{figure*}[tb]
  \centering
    \includegraphics[width=\textwidth]{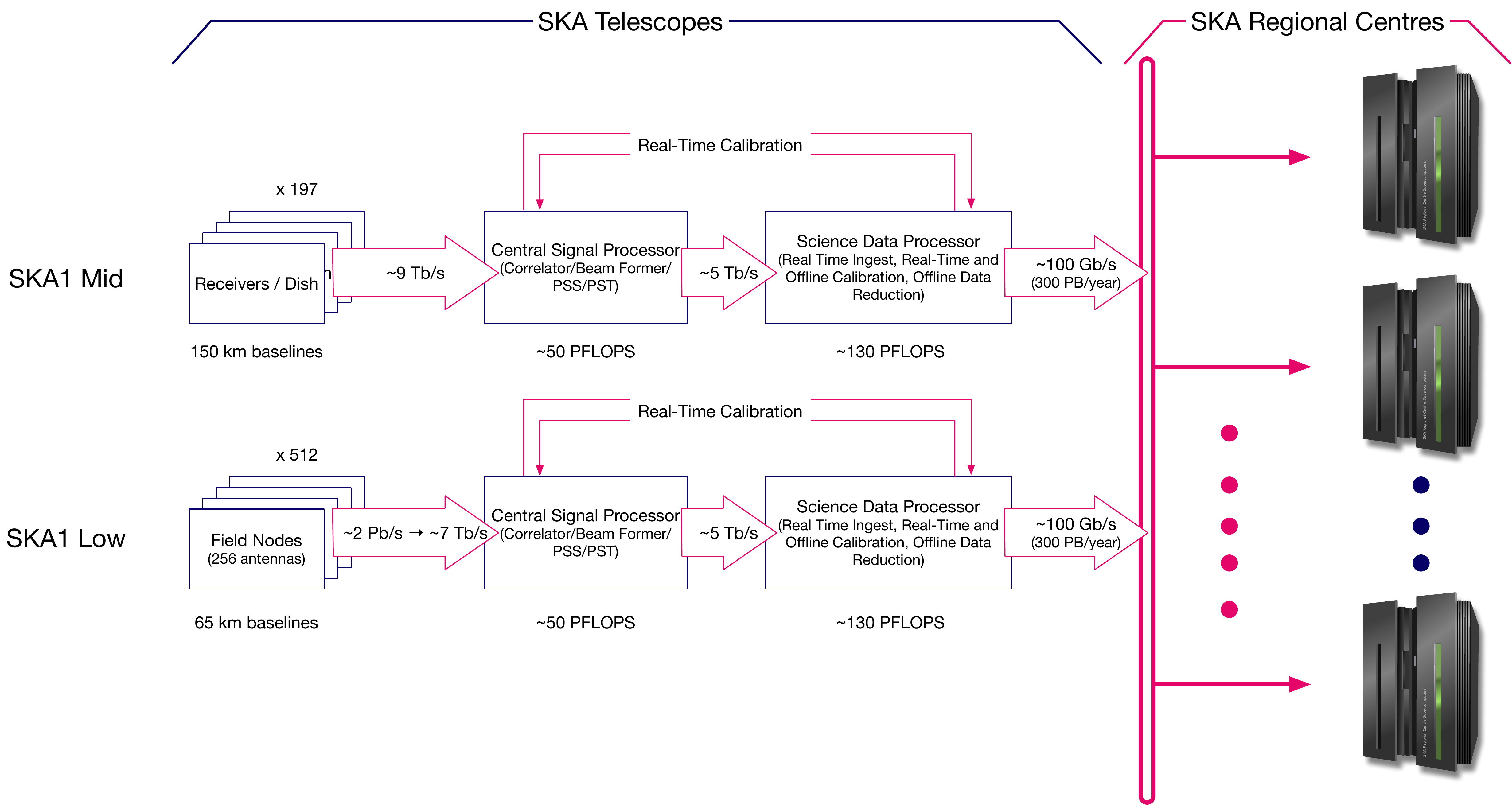}
  \caption{
      Notional data flow of the two SKA1 telescopes, SKA1-Mid and SKA1-Low. 
      The different maximum bandwidths for each flow are shown in each arrow.
      The data generated by the telescopes is distributed to the network of 
      SKA Regional Centres which are not in the scope of the SKA1 construction 
      project.
  }
  \label{fig:ska-data-flow}
\end{figure*}

Software plays a major role in the SKA project. A notional data flow view of the two telescopes is shown in Fig.~\ref{fig:ska-data-flow}. In there, some of the major blocks of the telescope requiring software are displayed:

\begin{itemize}
    \item The Science Data Processor for both telescopes includes the real-time ingest of the data being generated by the whole of the array. This is a big-data stream of up to 9 Terabits per second, and apart from the ingest of the bulk data and its relevant metadata, some real-time calibrations need to be executed in pseudo-realtime. Moreover, the system uses batch processing on those datasets to produce science-ready data for the network of SKA Regional Centres (SRCs) to proceed and deliver to our Principal Investigators and Co-Investigators.
    \item The SRCs themselves are supercomputing facilities that need to be able to perform High-Performance Computing analysis on demand from the community of SKA researchers.
    \item The Central Signal Processor are supercomputers in their own right, but instead of being general purpose they are mostly concerned with floating-point Multiply and Accumulate operations in support of Fast-Fourier Transform of the data, implemented as Field Programmable Gate Arrays (FPGAs). The firmware/bitimages that configure those FPGAs are also being managed through the same software processes as the rest of our software.    
\end{itemize}

In addition, two more large systems are software systems: the Telescope Management and Control (TMC) system is the control software that orchestrates and manages our telescopes, and is based on TANGO; and the Observatory and Science Operations (OSO) systems is the interface for researchers to submit observation proposals, and then convert them into the Scheduling Blocks that will be run by the TMC in order to carry out the observations.

\section{Current Status} 
\label{sec:current_status}

As indicated in Sec.~\nameref{sec:ska_to_skao}, the Construction phase has just started, but for software -- including the SKA telescopes' control system -- the development started in August 2018, using the bridging phase between pre-construction and construction to derisk our development processes and code. From that time until July 1st, 2021 we have gone already through ten full Program Increments, and the first PI to be partially in the construction phase was PI11, with PI12 -- our current PI as of the writing of this article -- being the first one fully within the SKA Phase 1 (SKA1) construction phase.

We are using the Scaled Agile Framework\footnote{\url{https://scaledagileframework.com}} (SAFe®) as the version of agile to be scaled to the needs of the SKA1 construction project. SAFe groups agile teams into Agile Release Trains (ARTs). Each ART is built around a stream of value to be delivered. Initially,  we started with a single train, which delivered all software required for the SKA telescopes. When the number of people and teams increased, we moved to two trains: one of them delivering the value for the Science Data Handling and Processing (SDH\&P) part, and another one for Observation Management and Controls (and Correlation; OMC). In PI11, we soft-launched a third ART, the Services ART, which provides value to both the DP and OMC ARTs in the form of System services (development platform, versioning, continous integration and continuous delivery), Platform services (definition and furnishing of the middleware that helps run all software, such as Kubernetes or Open Stack), and Networking services.

Key to the success of our software developments is providing a joined up Vision of what we want to deliver. And we update that Vision to the nearest horizon that is useful for our teams. There are 5 main milestones for the SKA telescope development, that we call Array Assemblies (AAs). In our initial roll-out plan, we had four AAs (AA1 to AA4). However, it was decided that an earlier integration testing point should be considered, and we call that milestone AA0.5.

Table~\ref{tab:array-assemblies} provides information on the different AA milestones, together with the capabilities that we expect to be able to test for each one, and when we currently expect to finish those milestones.

\begin{table}[!h]
   \centering
   \caption{Sensing elements per Array Assembly}
    \begin{tabular}{crr}
        \toprule
                           & \multicolumn{2}{c}{\textbf{Sensing Elements}}  \\
        \textbf{Milestone} & \multicolumn{1}{c}{\textbf{Mid}} & \multicolumn{1}{c}{\textbf{Low}}  \\
        \midrule
           AA0.5  &    4  &    6   \\
            AA1   &    8  &   16  \\
            AA2   &   64  &   64  \\
            AA3   &  121  &  256  \\
            AA4   &  197  &  512  \\
        \bottomrule
    \end{tabular}
   \label{tab:array-assemblies}
\end{table}

In the same vein of early testing, we started developing and planning the SKA software with our first Program Increment (PI1) in December 2018. The decision was to create a Minimum Viable Product (MVP) in the sense of Eric Rees~\cite{Rees:2009vi}: the minimum set of features that can solve a problem and that allows learning from early adopters to be incorported. In this case, our MVP is for the first integration prototype, and we named it the SKA MVP Product Integration, or SKAMPI.

An important part of SKAMPI is the support of the Continuous Integration/Continuous Deployment pipelines in order to test and verify our software. The work done so far is summarised by Matteo Di Carlo~\cite{TUBL04} in this conference.

\begin{figure*}[tb]
  \centering
    \includegraphics[width=\textwidth]{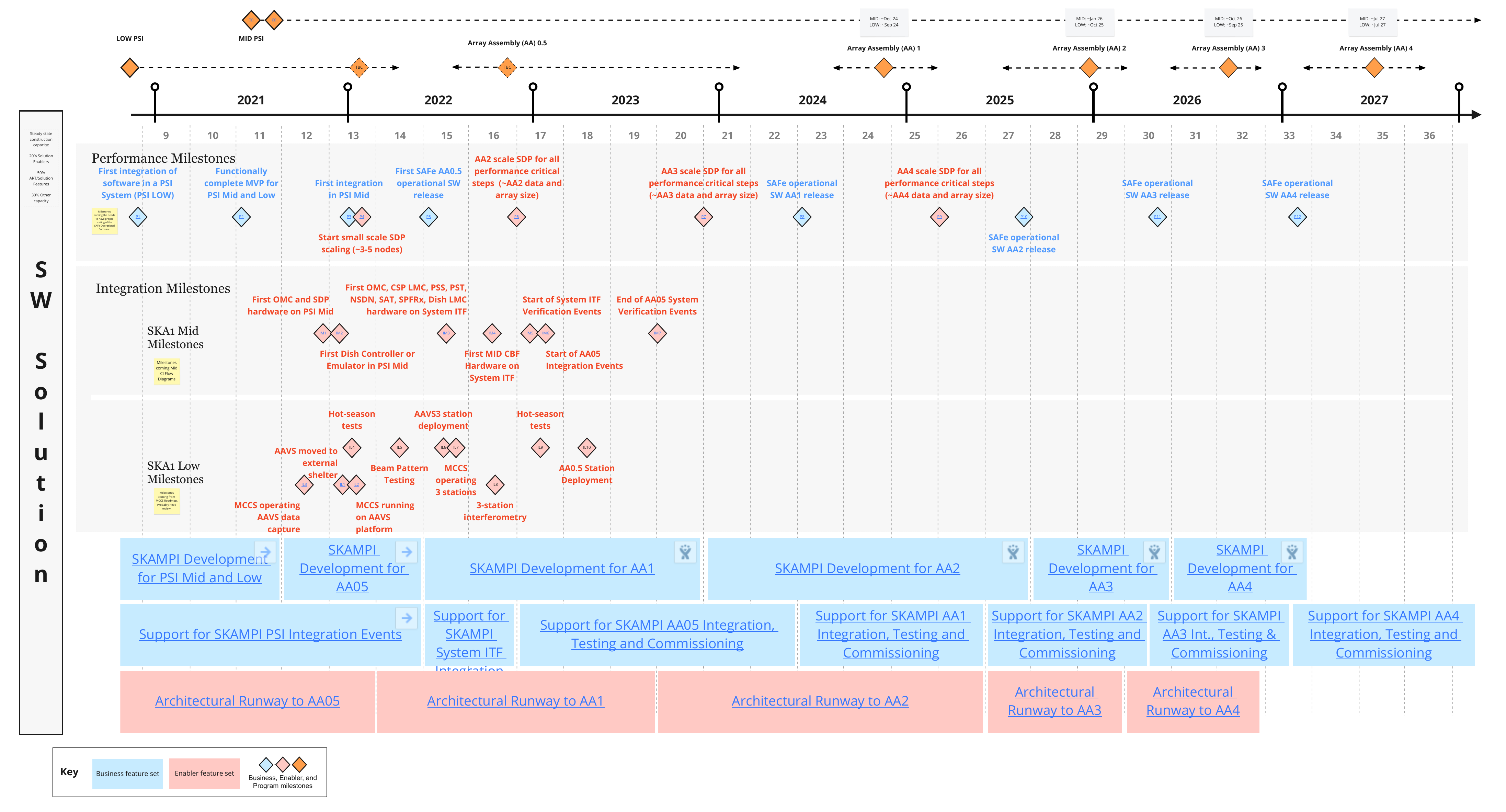}
  \caption{Long term Software Solution roadmap for all SKAO construction, and part of the pre-construction. Each Program Increment (PI) corresponds to 13 weeks, so there are exactly 4 PIs per year, and each PI planning event tends to happen within the same week every year, which helps with the predictability for the dates. The rhombs above the years corresponds to milestones that have to do with the telescope roll-out plans: PSIs stand for Prototype System Ingegrations. AA stands for Array Assembly, and each Array Assembly corresponds to increased number of equipment delivered, and additional capabilities. Performance Milestones have to do with when particular levels of performance, or non-functional requirements (NFRs), need to be met, in order to achieve the right level of performance when delivering the final system. Integration Milestones, on the other hand, indicate when particular functionality of the software needs to be available because of integration and testing requirements. The bottom row corresponds to three lanes of work: a) development stream of the SKA Minimum-viable-product Prototype Integration (SKAMPI); b) support stream of the usage of SKAMPI in several environments, including the AAs; and c) changes and evolution required in the software architecture related to the requirements of the different AAs.}
  \label{fig:solution-long-term-roadmap}
\end{figure*}

The development of SKAMPI in paraticular, and of all SKA software in general, needs to address the needs of each of the AAs indicated in Table~\ref{tab:array-assemblies}. To indicate what needs to be done in each moment, we have developed a Solution-level roadmap, which is updated at least every PI. And starting in PI11 we are also developing a joint TDT-level roadmap, that is then interpreted at software level. However, we have not yet managed to provide that joined up TDT/Solution-level roadmap view. The closest we have is the Solution level roadmap that can be seen in Fig.~\ref{fig:solution-long-term-roadmap}.

That roadmap follows three main development streams:

\begin{itemize}
    \item{\textbf{SKAMPI development:}} This stream includes all development effort required to provide working, useful SKAMPI software at each stage.
    \item{\textbf{SKAMPI integration support:}} This stream comprises development effort required to help the Assembly, Integration, and Verification (AIV) teams to test the SKA hardware and software.
    \item{\textbf{Architectural Runway:}} This final stream includes all development effort required to incrementally change the architecture of the software to address the needs of each of the individual AAs.
\end{itemize}

As indicated at the beginning of this section, we are currently executing PI12, and preparing the backlog and features to be planned in PI13, to be held in December 2021.


\section{Challenges} 
\label{sec:challenges}

During the pre-construction PIs, and in this first construction PI,  we have faced several challenges:

\begin{itemize}
    \item Using TANGO in an event-oriented way
    \item Containarisation and Orchestration
    \item Computing Scaling
    \item Team Scaling
\end{itemize}

\subsection{Using TANGO in an Event-Oriented Way} 
\label{sub:using_tango_in_an_event_oriented_way}

Most facilities using TANGO for control rely heavily on polling, as they control hardware systems for which it is expensive to obtain many of the monitoring values that might be required.

For SKA software, many of the controlled entities are other TANGO DeviceServers (DSs), and below those DSs additional software exists, which can easily change and/or generate values as required. In those cases, it is preferred to have an event-based system, in which data is only updated when an change-event is required.

Most SKA software has moved away from polling, and this has found some edge cases around the TANGO event mechanism, and the need for more TANGO expertise, and potentially the ability to contribute to change the TANGO event engine is required. 

There are additional issues coming from the interaction of the event system with containarisation and orchestration, but we will discuss those in the next subsection.

To try to resolve those issues, we have hired a new member of staff at SKAO that will contribute 50\% of their time to support the TANGO kernel, while providing additional TANGO development support to the teams. This is part of SKAO's contribution to the TANGO Collaboration, since we joined in October 2017.

We have also developed ways to log and trace events, as indicated by Samuel Twum~\cite{TUBL02} in this conference.


\subsection{Containarisation and Orchestration} 
\label{sub:containarisation_and_orchestration}

In order to be able to scale the execution and deployment of SKA software, and to orchestrate both its deployment and runtime environment, we are using Open Container Initiative (OCI) containers\footnote{url{https://opencontainers.org}} -- although we started with Docker\footnote{\url{http://docker.com}} containers. For orchestration, we started using Docker Compose\footnote{\url{https://docs.docker.com/compose/}}, but we have recetly moved to use Kubernetes\footnote{\url{http://kubernetes.io}} (K8s), and Helm\footnote{\url{https://helm.sh}} charts for managing all packaing, deployment, and software sequencing.

\begin{figure*}[tb]
  \centering
    \includegraphics[width=\textwidth]{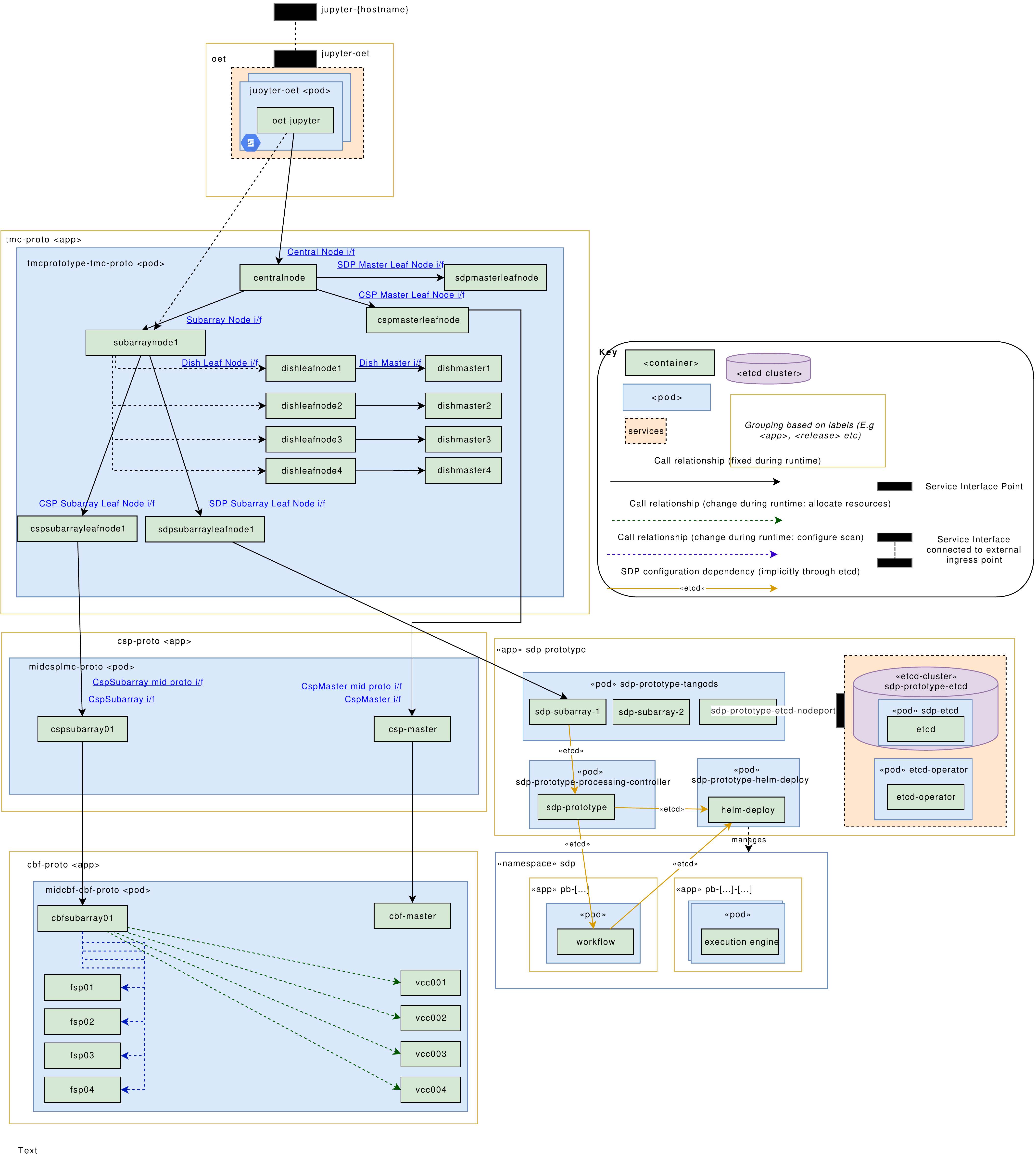}
  \caption{Connector and Component view of SKAMPI software. }
  \label{fig:skampi_context_diagram}
\end{figure*}

Figure~\ref{fig:skampi_context_diagram} shows the different labeled groupings of software, and how they are mapped into services, pods, and containers. Currently, a single TANGO DS leaves inside a container.

We are currently being challenged by the learning curve of K8s and Helm for some of our developers. This is another instance in which having an standardise approach within the TANGO community will be beneficial, and SKAO would like to foster such an outcome.


\subsection{Computing Scaling} 
\label{sub:computing_scaling}

As indicated in the previous subsection, we are currently using a 1:1 mapping between containers and TANGO DS, but also a 1:1 mapping between DSs and TANGO Devices. However, that granularity seems to impose a high-overhead, and we are looking into how to better combine DSs within containers, and Devices within Device Servers.

We are also finding some issues in how to express some of the dependencies, and linkages, through the Helm charts, to that we can properly map software releases, and combine them into a single product.


\subsection{Team Scaling} 
\label{sub:team_scaling}

The last challenge has to do with how our software development processes are scaling.

A large infrastructure project such as the construction of the SKA telescopes requires strong project management skills, and fixing a number of milestones to ensure proper monitoring of the progress. There is an Integrated Product Schedule (IPS) which is the source of truth for the major project milestones, and includes both schedule and costing information.

Converting that IPS into actionable information for the software development teams requires a lot of bandwidth between the TDTs and the software teams, and is something that we are working on to improve.

We are also starting to develop better rules for contribution and release management, that we expect to have in place soon.



\section{Next Steps} 
\label{sec:next_steps}

In order to address the previously outlined challenges, we are working to improve the communication and aligment between the IPS and the Software Roadmap: we start of TDT Planning as part of PI planning is helping with that, and we have also started to run dedicated roadmap alignment workshops to make sure that all dependencies and team needs are surfaced well ahead of PI planning.

We are also finishing with the qualification of suppliers, so that we can start proper contracting under construction funds, and finally move from bridging to construction funding.

Some of the issues with both Team Scaling and Computing Scaling have to do with lack of appropriate funding of effort to the System and Platform teams. We exected that setting up the contracting framework should help with this, by both accessing \emph{in-kind}, and commercial contributors to these teams.

Finally, we will keep iterating on our TANGO base software (as described by Sonja Vrcic~\cite{TUBR02} in this same conference), and keep iteraring our MVP of Product Integration in order to move to the first releases of proper SKAO Telescope Software.


\section{Conclusions} 
\label{sec:conclusion}

The construction phase that will deliver the transformational SKA telescopes has just started. After passing $\mathrm{T_{0}}$ in July 1st, 2021, we have already issued several contracts for construction ($\mathrm{C_{0}}$), but have also already been taking advantage of bridging funding in order to kickstart the development of the software. We have found some challenges, but we are doing so at the very start of the project, which makes it easier to resolve. We expect to be able to deliver our software on time, and to the satisfaction of our customers, which are primarily the AIV and Commissioning and Science Verification (CSV) teams.


\section{Acknowledgments} 
\label{sec:acknowledgments}

J. Santander-Vela wants to dedicate this paper to the memory of his recently passed-away father-in-law, Mr. Demetrio Sababa. He also wishes to acknowledge the work of all SKAO engineers, and the support of the co-authors.



\printbibliography

\end{document}